\documentstyle[amssymb,prl,aps,twocolumn,epsfig]{revtex} 

\newcommand {\beq} {\begin{equation}}
\newcommand {\eeq} {\end{equation}}
\newcommand {\beqa}{\begin{eqnarray}}
\newcommand {\eeqa}{\end{eqnarray}}
\newcommand {\del} {\partial}
\newcommand {\n}{\nonumber \\}
\newcommand {\tr}{{\rm tr\,}}
\newcommand {\Tr}{\mbox{Tr\,}}

\newcommand {\dd}{\mbox{d}}
\newcommand {\ee}{\mbox{e}}

\newcommand{\id}{{1\!\!1}} 

\font\mybb=msbm10 at 12pt
\def\bb#1{\hbox{\mybb#1}}

\def\IC{{\bb C}}
\def\IR{{\bb R}}

\begin{document}
\draft

\title{Exactly Solvable Matrix Models for the Dynamical\\
Generation of Space-Time in Superstring Theory}
\author{Jun Nishimura \cite{EmailJN}}
\address{The Niels Bohr Institute,
Blegdamsvej 17, DK-2100 Copenhagen \O, Denmark }
\date{preprint  NBI--HE--01--09, hep-th/0108070; \today}
\twocolumn[\hsize\textwidth\columnwidth\hsize\csname@twocolumnfalse\endcsname

\maketitle
\begin{abstract}
We present a class of solvable SO($D$) symmetric matrix models
with $D$ bosonic matrices coupled to chiral fermions.
The SO($D$) symmetry is spontaneously broken
due to the phase of the fermion integral.
This demonstrates the conjectured mechanism for the 
dynamical generation of four-dimensional space-time in the IIB matrix model,
which was proposed as a nonperturbative definition of 
type IIB superstring theory in ten dimensions.
%
%
\end{abstract}



]   

\paragraph*{Introduction.---}

One of the biggest puzzles in superstring theory 
is that the space-time dimensionality
which naturally allows a consistent construction of the theory is ten
instead of four.
A natural resolution of this puzzle is
to consider our 4d space time to appear {\em dynamically} 
and the other 6 dimensions to become invisible 
due to some nonperturbative effects.
This may be compared to the situation with QCD in the early 70's,
where quarks were introduced to explain high-energy experiments,
while the puzzle was that none of them has ever been observed
in reality. 
The understanding of quark confinement as 
a nonperturbative phenomenon in non-Abelian gauge theories 
was important for QCD to
be recognized as the correct theory of strong interaction.
Likewise, we think it important to try to understand 
the puzzle of space-time dimensionality
in terms of nonperturbative dynamics of superstring theory.
Nonperturbative formulations
of superstring/M theories proposed in Ref.~\cite{BFSS,IKKT} 
may play an analogous role as the lattice gauge theory.

The issue of the dynamical generation of space-time has been
pursued \cite{AIKKT,HNT,NV,branched,Burda:2000mn,4dSSB,sign} 
in the context of the IIB matrix model \cite{IKKT},
which was proposed as a nonperturbative definition
of type IIB superstring theory in ten dimensions.
This model is a supersymmetric matrix model
composed of 10 bosonic matrices 
and 16 fermionic matrices,
and it can be thought of as the zero-volume limit of
10d SU($N$) super Yang-Mills theory \cite{endnote0}.
The 10 bosonic matrices represent the dynamical
space-time \cite{AIKKT} and the model has manifest SO(10) symmetry.
Our four-dimensional space-time
may be accounted for, 
if configurations with only four extended 
directions dominate the integration over 
the bosonic matrices. 
This in particular requires that the SO(10) symmetry to be spontaneously
broken. 
It is suggested that the phase of the fermion integral
plays a crucial role 
in such a phenomenon \cite{NV,branched,sign}.

In this Letter, we present a class of solvable 
SO($D$) symmetric matrix models,
where $D$ bosonic hermitian matrices are interpreted
as the space-time.
The models also include chiral fermions which yield
complex fermion integrals as in the IIB matrix model.
We study the $D=4$ case explicitly and find that 
the SO(4) symmetry is broken down to SO(3);
namely 3 dimensional space-time
is generated dynamically in the SO(4) invariant model.
If we replace the fermion integral by its absolute value,
the model is still solvable and exhibits no SSB.
This result clearly demonstrates the conjectured mechanism
for the dynamical generation of space-time
in the IIB matrix model.

\paragraph*{The model.---}
The partition function of the model we consider 
is given by
\beqa
Z &=&  \int \dd A \dd \psi \dd \bar{\psi} ~ 
\ee^{-(S_{\rm b} + S_{\psi})}  \ ,
\label{original_model} \\
S_{\rm b} &=& \frac{1}{2}N \, \tr( A_{\mu}^2) 
\label{actionB}
\\ 
S_{\psi} &=& - \bar{\psi}_\alpha^f (\Gamma_\mu)_{\alpha\beta}
A_\mu \psi_\beta^f \ ,
\label{actionPSI}
\eeqa
where $A_{\mu}$ ($\mu = 1,\cdots , D$) are $N\times N$ hermitian
matrices
and $\bar{\psi}_\alpha^f$, $\psi_\alpha^f$ are $N$-dimensional
row and column vectors, respectively.
(The system has an SU($N$) symmetry.)
We assume that $D$ is even, but we comment on a generalization to
odd $D$ later.
The actions (\ref{actionB}) and (\ref{actionPSI}) are 
SO($D$) invariant,
where the bosonic matrices $A_\mu$ transform as a vector,
and the fermion fields $\bar{\psi}_\alpha^f$ and $\psi_\alpha^f$ 
transform as Weyl spinors.
The spinor index $\alpha$ runs over $1, \cdots , p$, 
where $p= 2 ^{D/2-1}$ is the dimension of the spinor space.
The $p\times p$ matrices $\Gamma_\mu$ 
are the gamma matrices after Weyl projection.
The flavour index $f$ runs over $1, \cdots , N_f$.
We take the large $N$ limit with $r \equiv N_f / N $ fixed
(Veneziano limit) \cite{endnote5}.
The fermionic part of the model can be thought of 
as the zero-volume limit of the system of 
Weyl
fermions in $D$-dimensions interacting with a background gauge field
via fundamental coupling.

Integrating out the fermion fields, one obtains
\beq
Z =  \int \dd A \, \ee^{-S_{\rm b} } (\det {\cal D})^{N_f} \ ,
\label{det_model}
\eeq
where ${\cal D}$ is a $p N \times p N$ matrix given by
${\cal D} = \Gamma_\mu A_\mu$.
In $D=2$, we find that $\det {\cal D}$ transforms under
an SO(2) transformation as 
$\det {\cal D} \mapsto \ee ^{i\theta} \det {\cal D}$,
where $\theta$ is the angle of rotation \cite{endnote1}.
Hence the partition function (\ref{det_model}) vanishes
in this case \cite{endnote2}.
In $D\ge 4$, $\det {\cal D}$ is SO($D$) invariant and so 
is the model. 

The fermion determinant $\det {\cal D}$ is complex in general.
Under parity transformation : 
$A^P _1  = - A_D$ and $A^P_i = A_i$ 
for ($i \neq D$),
it 
becomes complex conjugate.
From this, it follows that $\det {\cal D}$ becomes real if 
$A_D = 0$, or more generally, if $n_\mu A_\mu = 0$ for 
some vector $n_\mu$.

We interpret the $D$ bosonic $N \times N$ hermitian matrices
$A_{\mu}$ as the dynamical space-time 
as in the IIB matrix model \cite{AIKKT}.
The space-time has the Euclidean signature 
as a result of the Wick rotation, which is always necessary
in path-integral formalisms.
In the present model, we can obtain the extent of space time
\beq
R^2 \equiv \left\langle \frac{1}{N} \tr (A_\mu)^2 \right\rangle =
D + r p 
\label{resultR}
\eeq
using a scaling argument for arbitrary $N$. 

In order to probe the possible SSB of SO($D$),
we first generalize the bosonic action as
\beq
S_{\rm b}(\vec{m})
= \frac{1}{2}N \, \sum_{\mu} m_\mu \tr( A_{\mu}^2) \ . 
\label{anisotropic}
\eeq
We calculate the extent in the $\mu$-th direction
\beqa
\lambda_\mu &=& \left\langle \frac{1}{N} \tr (A_\mu)^2  
\right\rangle_{\vec{m}}
\mbox{~~~(no summation over $\mu$)} \n
&=& - \frac{2}{N^2} \frac{\del}{\del m_\mu} \ln Z(\vec{m}) 
\label{deriv}
\eeqa
for arbitrary $\vec{m}$ in the large $N$ limit.
Then we take the limit of $m_\nu \rightarrow 1$ 
(for all $\nu$) keeping the order 
\beq
m_1 < m_2 < \cdots < m_D \ .
\label{order}
\eeq
If $\lambda_\mu$ do not converge to the same value, 
it signals the SSB of SO($D$) symmetry.

\paragraph*{The method.---}
The model (\ref{original_model}) with the anisotropic bosonic action
(\ref{anisotropic})
can be solved in the large $N$ limit
by using a technique known from Random Matrix Theory \cite{RMT}.
Integrating out the bosonic matrices $A_\mu$,
\beqa
&~&Z \sim \frac{1}{{\cal N}}
\int \dd \psi \dd \bar{\psi} 
\exp \Bigl(- \frac{1}{2N} S_{\rm Fermi} \Bigr) \ ,
\label{integrate_out}  \\
&~&S_{\rm Fermi} =  
(\bar{\psi}_\alpha^f \psi_\beta^g )
\Sigma _{\alpha\beta,\gamma\delta}
(\bar{\psi}_\gamma^g \psi_\delta^f )
\label{four-fermi} \\
&~&\Sigma _{\alpha\beta,\gamma\delta}
= \sum_{\mu} \frac{1}{m_\mu} (\Gamma_\mu)_{\alpha\delta}
(\Gamma_\mu)_{\gamma\beta} \ .
\eeqa
The normalization factor ${\cal N}$ in (\ref{integrate_out})
is given
by 
\beq
{\cal N} =  \prod _\mu (m_\mu)^{N^2/2}  \ .
\label{normalization}
\eeq
Here and henceforth, we omit irrelevant $\vec{m}$-independent factors
in the partition function.

The four-fermi action (\ref{four-fermi}) can be written as
\beqa
S_{\rm Fermi} &=&  \Sigma_{\alpha\beta,\gamma\delta} 
\Bigl( \Phi _{\alpha\beta,fg} ^{(+)}
\Phi _{\gamma\delta,fg} ^{(+)} 
-  \Phi _{\alpha\beta,fg} ^{(-)} \Phi _{\gamma\delta,fg} ^{(-)} \Bigr) \ , 
\label{four-fermi2}
\\
&~&\Phi_{\alpha\beta,fg} ^{(+)} = \frac{1}{2}
(\bar{\psi}_\alpha^f \psi_\beta^g 
+ \bar{\psi}_\alpha^g \psi_\beta^f ) \n
&~& \Phi_{\alpha\beta,fg} ^{(-)} = \frac{1}{2} 
(\bar{\psi}_\alpha^f \psi_\beta^g 
- \bar{\psi}_\alpha^g \psi_\beta^f )  \ .
\eeqa
The matrix $\Sigma$, where we consider ($\alpha\beta$)
and ($\gamma\delta$) as single indices,
is symmetric, and one can always make it real 
by choosing the representation of $\Gamma_\mu$ properly.
Hence one can diagonalize it as
\beq
\Sigma _{\alpha\beta,\gamma\delta}
= \sum _{\rho\tau} 
O_{\alpha\beta , \rho \tau} \Lambda_{\rho\tau} 
O_{\gamma\delta , \rho \tau } \ ,
\label{Sigmadiag}
\eeq
and (\ref{four-fermi2}) can be written as
\beqa
S_{\rm Fermi} &=&  
\sum_{\rho\tau}
\Lambda_{\rho\tau} 
\Bigl( \sum_{\alpha\beta} 
O_{\alpha\beta , \rho \tau} \Phi_{\alpha\beta,fg}^{(+)} \Bigr)^2  \n
&~&  
- \sum_{\rho\tau}
\Lambda_{\rho\tau} 
\Bigl( \sum_{\alpha\beta} 
O_{\alpha\beta , \rho \tau} \Phi_{\alpha\beta,fg}^{(-)} \Bigr)^2  \ .
\label{diag4}
\eeqa

Each square in eq.~(\ref{diag4}) 
can be linearized by a Hubbard-Stratonovitch transformation 
according to 
\beq
\exp(-A Q^2) \sim 
\int \dd \sigma \exp
\Bigl(- \frac{\sigma^2}{ 4A} - i Q \sigma \Bigr) \ .
\eeq
Introducing $p^2$ complex matrices
$\hat{\sigma}_{\rho\tau}$ of size $N_f \ $, we arrive at
\beqa
Z &\sim& \frac{1}{{\cal N}} \int \dd \hat{\sigma}  \dd \psi \dd \bar{\psi} 
\exp ( - N S_G + S_Q ) \\
&~&S_{\rm G}=\Tr (\hat{\sigma}_{\rho\tau}^\dag
\hat{\sigma}_{\rho\tau}) ~~~;~~~
S_{\rm Q} =  \bar{\psi}_{\alpha}^f  {\cal M}_{\alpha\beta}^{fg}
\psi_{\beta}^g    \ ,
\eeqa
where the $p \, N_f \times p \, N_f$ matrix ${\cal M}$ is
\beq
{\cal M}_{\alpha \beta}^{fg}
= \frac{1}{\sqrt{2}}\sum_{\rho\tau}
\sqrt{\Lambda_{\rho\tau}} O_{\alpha\beta , \rho\tau} 
(\hat{\sigma}_{\rho\tau} + \hat{\sigma}_{\rho\tau}^\dag)_{fg}  \ .
\eeq
The fermionic integration yields
\beq
Z \sim \frac{1}{{\cal N}} \int \dd \hat{\sigma} 
\exp (- N W[\hat{\sigma}]) \ ,
\eeq
where the effective action $W[\hat{\sigma}]$ is given by
\beq
W[\hat{\sigma}] = S_{\rm G} - \ln \det {\cal M} \ .
\eeq

In the large $N$ limit, 
the evaluation of the partition function amounts to
solving the saddle-point equations, which are given by
\beq
(\hat{\sigma}_{\rho\tau})_{fg} = (\hat{\sigma}_{\rho\tau}^\dag)_{fg}
= \frac{1}{\sqrt{2}} \sum_{\alpha\beta} ({\cal M}^{-1})_{\beta\alpha}^{fg} 
\sqrt{\Lambda_{\rho\tau}}
O_{\alpha\beta,\rho\tau}   \ .
\label{spa0}
\eeq
Assuming that the flavour SU($N_f$) symmetry is not broken,
we set $\hat{\sigma}_{\rho\tau}= \sigma_{\rho\tau} \id$,
where $\sigma_{\rho\tau} \in \IC $.
We can further take $\sigma_{\rho\tau}$ to be real, due to 
(\ref{spa0}).
Then the effective action reduces to
\beq
W = N_f \Bigl\{ (\sigma_{\rho\tau})^2 - \ln \det M  (\sigma)\Bigr\} \ , 
\label{spag_reduced}
\eeq
where the $p \times p$ matrix $M (\sigma)$ is given by
\beq
M_{\alpha \beta} (\sigma)
= \sqrt{2} \sum_{\rho\tau}
\sqrt{\Lambda_{\rho\tau}} O_{\alpha\beta , \rho\tau} 
\sigma_{\rho\tau}  \ .
\label{defMgen}
\eeq
Thus the problem reduces to a system of finite degrees of freedom.

\paragraph*{Exact results in 4d.---}
Let us solve the saddle-point equations explicitly
in the simplest case $D=4$.
We choose $\Gamma_i$ ($i=1,2,3$) to be Pauli matrices
and $\Gamma _4 = i \id$.
The matrix $M$ is a $2 \times 2 $ matrix
\beqa
&~& M (\sigma) = 
\left( 
\begin{array}{cc}
a + i b  & 
i c +  d  \\
i c -  d & 
a - i b
\end{array}
\right) \ ,
\label{defM} \\
&~& a = \sqrt{\rho_4} \, \sigma_{11} 
~~~;~~~b = \sqrt{\rho_3} \, \sigma_{22} 
 \n
&~& c = \sqrt{\rho_1} \, \sigma_{12} 
~~~;~~~d = \sqrt{\rho_2} \, \sigma_{21}
 \ ,
\eeqa
where we have introduced the notation
\beq
\rho_\mu = \sum_{\nu} (-1)^{\delta_{\mu\nu}} (m_\nu)^{-1}  \ .
\eeq
The saddle-point equations are
\beqa
\sigma_{11} =  \Delta ^{-1}  \rho_4 \, \sigma_{11} 
~~~&;&~~~
\sigma_{12} =  \Delta ^{-1}  \rho _1  \, \sigma_{12} \n
\sigma_{21} =  \Delta ^{-1} \rho_2 \, \sigma_{21} 
~~~&;&~~~
\sigma_{22} =  \Delta ^{-1} \rho_3 \, \sigma_{22}  \ ,
\label{spa}
\eeqa
where $\Delta = a^2 + b^2 + c^2 + d^2$.
Eq.~(\ref{spa}) implies that $\Delta$ should 
take one of the four possible values
$ \rho _1$, $ \rho_2$, $ \rho_3$ and $ \rho_4$.
In each case, the effective action
is evaluated as $W =  N_f ( 1 - \ln \Delta )\ $. 

When the parameters $\vec{m}$ obey the order (\ref{order}),
the dominant saddle-point is given by $\Delta =  \rho_4$.
Thus the partition function can be obtained as
\beq
Z \sim \frac{1}{{\cal N}} \,
\ee ^{N N_f \ln  \rho_4} \ .
\eeq
Using (\ref{deriv}) we get
\beq
\lambda_\mu =
(m_\mu)^{-1} \pm 2 r \frac{1}{\rho_4}(m_\mu)^{-2} \ ,
\label{exactresult}
\eeq
where the $\pm$ symbol should be $+$ for $\mu = 1,2,3$ and
$-$ for $\mu = 4$.
In the limit of $m_\nu \rightarrow 1$ (for all $\nu$), one obtains
\beq
\lambda_1 = \lambda_2 = \lambda_3 = 1 +  r \mbox{~~~};\mbox{~~~}
\lambda_4 = 1 - r \ ,
\label{SSBresult}
\eeq
which means that the SO(4) is spontaneously broken down to SO(3).
We note that $R^2 = \sum _\mu \lambda_\mu = 4 + 2r$ 
agrees with the finite $N$ result (\ref{resultR}).
The SSB is associated with the formation of a condensate
$\langle \bar{\psi}_\alpha ^f \psi_\alpha ^f  \rangle$, which
is invariant under SO(3), but not under full SO(4).

\paragraph*{The phase of the determinant.---}
In order to clarify the role played by the
phase of the determinant $\det {\cal D}$,
let us consider the model
\beq
Z' =  \int \dd A \, \ee^{-S_{\rm b} } |\det {\cal D}|^{N_f} \ . 
\label{modified_model_det}
\eeq
This model can be obtained by replacing half of the 
$N_f$ Weyl fermions $\psi$ in (\ref{original_model}) by
Weyl fermions $\chi$ with opposite chirality.
Namely, eq.\ (\ref{modified_model_det})
can be rewritten as
\beqa
Z' &=&  \int \dd A \dd \psi \dd \bar{\psi} \dd \chi \dd \bar{\chi} ~ 
\ee^{-(S_{\rm b} + S_{\psi} + S_{\chi})}  \ ,
\label{modified_model}\\
S_{\chi} &=& - \bar{\chi}_\alpha^f (\Gamma_\mu ^\dag)_{\alpha\beta}
A_\mu \chi_\beta^f \ .
\eeqa
We use a representation of gamma matrices in which
$\Gamma_i$ ($i=1,\cdots , (D-1)$) are hermitian and $\Gamma_D = i \id$.
Note that
the flavour index $f$ now runs over $f = 1, \cdots , N_f /2$.

We can solve the above model
with the anisotropic bosonic action
(\ref{anisotropic}) in the large $N$ limit using the same method
as before.
The four-fermi action reads
\beqa
&~&S'_{\rm Fermi} =  
(\bar{\psi}_\alpha^f \psi_\beta^g )
\Sigma _{\alpha\beta,\gamma\delta}
(\bar{\psi}_\gamma^g \psi_\delta^f ) 
+ (\bar{\chi}_\alpha^f \chi_\beta^g )
\Sigma _{\alpha\beta,\gamma\delta}
(\bar{\chi}_\gamma^g \chi_\delta^f ) \n
&~&\mbox{~}+ (\bar{\psi}_\alpha^f \chi_\beta^g )
\widetilde{\Sigma} _{\alpha\beta,\gamma\delta}
(\bar{\chi}_\gamma^g \psi_\delta^f ) 
+ (\bar{\chi}_\alpha^f \psi_\beta^g )
\widetilde{\Sigma} _{\alpha\beta,\gamma\delta}
(\bar{\psi}_\gamma^g \chi_\delta^f )  \ ,
\label{four-fermi3}
\eeqa
where $\widetilde{\Sigma}$ can be obtained from $\Sigma$ 
by replacing $m_D$ by $- m_D$.
Similarly to (\ref{Sigmadiag}), it can be diagonalized as
\beq
\widetilde{\Sigma} _{\alpha\beta,\gamma\delta}
= \sum _{\rho\tau} 
\widetilde{O}_{\alpha\beta , \rho \tau} \widetilde{\Lambda}_{\rho\tau} 
\widetilde{O}_{\gamma\delta , \rho \tau } \ .
\eeq
In order to linearize (\ref{four-fermi3}), we have to 
introduce four sets of $\hat{\sigma}_{\rho\tau}$ matrices,
which we denote as $\hat{\sigma}_{\rho\tau}^{\psi}$,
$\hat{\sigma}_{\rho\tau}^{\chi}$,
$\hat{\sigma}_{\rho\tau}^{S}$ and $\hat{\sigma}_{\rho\tau}^{A} \ $.
As before, we set 
$\hat{\sigma}_{\rho\tau}^\psi= \sigma_{\rho\tau}^\psi \id$,
where $\sigma_{\rho\tau} \in \IR $, etc..
Introducing a new complex variable $\tilde{\sigma}_{\rho\tau} 
=\frac{1}{\sqrt{2}} (\sigma_{\rho\tau}^{S} + i \sigma_{\rho\tau}^{A})$,
the effective action becomes
\beqa
W' &=& \frac{N_f}{2} ( \, S'_{\rm G} - \ln \det M' ) \ , \\
S'_{\rm G}&=&  (\sigma_{\rho\tau}^\psi)^2
+ (\sigma_{\rho\tau}^\chi)^2
+ 2 \, | \tilde{\sigma}_{\rho\tau} | ^2   \\
M' &=& \left(
\begin{array}{cc}
M(\sigma^\psi ) & 
\widetilde{M}(\tilde{\sigma})  \\
\widetilde{M}(\tilde{\sigma}^* )
 & M(\sigma^\chi ) 
\end{array}
\right)  \ .
\eeqa
The $p \times p$ matrices $M(\sigma^\psi )$ and $M(\sigma^\chi)$
are the same as (\ref{defMgen})
except that $\sigma_{\rho\tau}$ is replaced by
$\sigma_{\rho\tau} ^{\psi}$ and $\sigma_{\rho\tau} ^{\chi}$,
respectively.
The new $p \times p$ matrix $\widetilde{M}(\tilde{\sigma})$ is given by
\beq
\widetilde{M}_{\alpha \beta}(\tilde{\sigma})
= \sqrt{2} \sum_{\rho\tau}
\sqrt{\widetilde{\Lambda}_{\rho\tau}} 
\widetilde{O}_{\alpha\beta , \rho\tau} \tilde{\sigma}_{\rho\tau} \ .
\eeq
The set of solutions to the saddle-point equations
is richer than before. There are solutions with 
$\tilde{\sigma}_{\rho\tau} = \tilde{\sigma} _{\rho\tau} ^{*} = 0 $.
In this case, the problem reduces to the previous one.
However, there is another class of solutions in which 
$\sigma _{\rho\tau} ^{\psi} = \sigma _{\rho\tau} ^{\chi} = 0 $.

Let us consider the $D=4$ case.
The matrix $\widetilde{M}$ is 
a $2 \times 2 $ matrix
\beqa
&~& \widetilde{M}(\tilde{\sigma}) = 
\left( 
\begin{array}{cc}
\tilde{a} + i \tilde{b}  & i \tilde{c} +  \tilde{d}  \\
i \tilde{c} -  \tilde{d} & \tilde{a} - i \tilde{b}
\end{array}
\right) \ ,
\label{defMprime} \\
&~& \tilde{a} = \sqrt{\rho} \, \tilde{\sigma}_{11}
~~~;~~~\tilde{b} = \sqrt{\rho_{34}} \, \tilde{\sigma}_{22} \n
&~&\tilde{c} = \sqrt{\rho_{14}} \, \tilde{\sigma}_{12}
~~~;~~~\tilde{d} = \sqrt{\rho_{24}} \, \tilde{\sigma}_{21} \ ,
\eeqa
where we have introduced the notations
\beq
\rho = \sum_{\nu}  (m_\nu)^{-1} ~;~
\rho_{\mu\lambda} = \sum_{\nu} 
(-1)^{\delta_{\mu\nu} + \delta_{\lambda\nu}} (m_\nu)^{-1}  \ .
\eeq

For the first class of solutions, 
the effective action at each saddle-point is given by
$W '  = N_f ( 1 - \ln \rho_\nu ) \ $, where $\nu = 1,2,3,4$.
For the second class of solutions, 
the saddle-point equations become
\beqa
\tilde{\sigma}_{11} ^* =  \widetilde{\Delta} ^{-1}
  \rho \, \tilde{\sigma}_{11}
~~~&;&~~~
\tilde{\sigma}_{12} ^*=  \widetilde{\Delta} ^{-1}  
\rho _{14}  \, \tilde{\sigma}_{12}  \n
\tilde{\sigma}_{21} ^*  =  \widetilde{\Delta} ^{-1} 
\rho_{24} \, \tilde{\sigma}_{21}
~~~&;&~~~
\tilde{\sigma}_{22} ^* =  \widetilde{\Delta} ^{-1} 
\rho_{34} \, \tilde{\sigma}_{22}   \ ,
\label{spa2}
\eeqa
and their complex conjugates,
where $\widetilde{\Delta} = \tilde{a}^2 +
\tilde{b}^2 + \tilde{c}^2 + \tilde{d}^2 $.
Due to (\ref{spa2}),
$|\widetilde{\Delta}|$ should take one of the four values
$\rho $, $\rho_{14}$, $\rho_{24}$ and $\rho_{34}$.
In each case, the effective action
is evaluated as 
$W' =  N_f ( 1 - \ln |\widetilde{\Delta}|) \ $.

Thus for arbitrary $\vec{m}$,
we find that the dominant saddle-point is 
given by the second class of the solutions 
with $|\widetilde{\Delta}| = \rho $
and the partition function is obtained as
\beq
Z' \sim \frac{1}{{\cal N}} \,
\ee ^{N N_f \ln  \rho} \ .
\eeq
Using (\ref{deriv}) we get
\beq
\lambda_\mu =
(m_\mu)^{-1} + 2 r \frac{1}{\rho} (m_\mu)^{-2}
\rightarrow 1 +  \frac{1}{2} r  \ ,
\eeq
in the limit of $m_\nu \rightarrow 1$ (for all $\nu$),
which means that SO(4) is preserved.
A nonvanishing condensate $\langle 
\bar{\psi}_\alpha ^f \chi_\alpha ^f 
+ \bar{\chi}_\alpha ^f \psi_\alpha ^f 
\rangle$ breaks chiral symmetry, but not SO(4).

\paragraph*{Discussion.---}
An interesting feature of the exact result (\ref{SSBresult})
is that $\lambda_4$ decreases linearly as $r\equiv N_f/N$ is increased.
At $r=1$ \cite{endnote6}, 
$\lambda_4$ becomes zero, and the dynamical space-time 
becomes completely 3-dimensional.
When $r > 1$, $\lambda_4$ becomes negative. 
This is possible because the VEV of a real positive
observable is not necessarily real positive
if the weight involves a phase.

One can generalize the model to odd $D$ by considering 
Dirac fermions instead of Weyl fermions.
In fact, such a model can be obtained from the even $D$ model
considered here by taking the $m_D \rightarrow \infty$ limit.
The result for the 3d case can thus
be read off from (\ref{exactresult})
as $\lambda _\mu = 1 + \frac{2}{3}r$ for all $\mu$,
which preserves the SO(3) symmetry.
We note that in the odd $D$ models the fermion determinant 
for each flavour is real, 
but it is not necessarily positive.
However, for even $N_f$ one obtains a real positive weight,
and for odd $N_f$ the sign of the weight is independent of $N_f \ $.
This explains the absence of SSB in the 3d model.

We think that our analytical results put the conjectured mechanism
for the dynamical generation of space-time
on firmer ground.
The phase of the fermion integral favours lower dimensional configurations
and as a result the space-time collapses in the large $N$ limit.
On the other hand, the actual dimensionality of the dynamical space-time
in the IIB matrix model is yet to be determined.
Unfortunately standard Monte Carlo simulation is difficult
precisely due to the existence of the phase.
However, we have recently proposed a new method to circumvent 
this problem \cite{sign}.
Analytical approaches 
using approximations such as the one in Ref.\ \cite{Sugino:2001fn} 
may also be useful.
We hope that the models presented in this Letter will serve also as a
testing ground for new ideas to understand this interesting phenomenon.

\paragraph*{Acknowledgments.---}
The author would like to thank
H.\ Kawai, F.\ Sugino and G.\ Vernizzi for discussions,
which motivated this work.
He is also grateful to J.J.\ Verbaarschot
for correspondence on exact results in Random Matrix Theory.


\begin{references}
\bibitem{endnote5} When $N_f = N$, 
the fermion fields $\psi_{\alpha}^f$ and $\bar{\psi}_{\alpha}^f$ 
can be written in terms of $N \times N$
fermionic matrices $(\Psi_\alpha)_{i f}$ and $(\bar{\Psi}_\alpha)_{f i}$,
so that the fermion action reads $S_\psi = - \tr (\bar{\Psi}_\alpha
\Gamma_\mu A_\mu \Psi_\alpha )\ $.
\bibitem{endnote2} 
If we use the anisotropic bosonic action (\ref{anisotropic}),
the partition function of the $D=2$ model can be calculated
as $Z \sim {\cal N}^{-1} \{(m_1)^{-1} - (m_2)^{-1} \}^{r N^2/2}$ 
in the large $N$ limit, where ${\cal N}$ is given by (\ref{normalization}).
\bibitem{endnote6}
According to \cite{endnote5},
$r = 1$ is the case in which the 4d model becomes
similar to the 4d version of the IIB matrix model concerning
the number of degrees of freedom.
\end{references}
\end{document}